\documentclass[conference,10pt]{IEEEtran}
\ifCLASSINFOpdf
\usepackage[pdftex]{graphicx}
\graphicspath{{../pdf/}{../jpeg/}}
\DeclareGraphicsExtensions{.pdf,.jpeg,.png}
\else
\usepackage[dvips]{graphicx}
\graphicspath{{../eps/}}
\DeclareGraphicsExtensions{.eps}
\fi
\usepackage[cmex10]{amsmath}
\newcommand{\tr}{\tilde{R}}
\newcommand{\mt}{\mathcal{T}}

\newcommand{\hm}{\hat{m}}

\newcommand{\mc}{\mathcal{C}}

\usepackage{mdwmath}
\usepackage{mdwtab}
\newtheorem{theorem}{Theorem}
\newtheorem{definition}{Definition}

\IEEEoverridecommandlockouts
\begin{document}
\title{Key Agreement Over A State-Dependent 3-Receiver Broadcast Channel
}
%\author{\IEEEauthorblockN{Michael Shell}
%\IEEEauthorblockA{School of Electrical and\\Computer Engineering\\
%Georgia Institute of Technology\\
%Atlanta, Georgia 30332--0250\\
%Email: http://www.michaelshell.org/contact.html}
%\and
\author{\IEEEauthorblockN{Mohsen Bahrami, Ali Bereyhi, Sadaf Salehkalaibar and Mohammad Reza Aref \thanks{This work was partially supported by Iranian NSF under contract no. $88114/46-2010$.}}
\IEEEauthorblockA{Information Systems and Security Lab (ISSL),\\
Sharif University of Technology, Tehran, Iran,\\
Email: \{bahramy, bereyhi, s{\_}saleh\}@ee.sharif.edu, aref@sharif.edu }
}
%\and
%\IEEEauthorblockN{James Kirk\\ and Montgomery Scott}
%\IEEEauthorblockA{Starfleet Academy\\
%San Francisco, California 96678-2391\\


%Telephone: (800) 555--1212, Fax: (888) 555--1212}


\maketitle

\begin{abstract}
%\boldmath
In this paper, we consider the problem of secret key agreement in state-dependent 3-receiver broadcast channels. In the proposed model, there are two legitimate receivers, an eavesdropper and a transmitter where the channel state information is non-causally available at the transmitter. We consider two setups. In the first setup, the transmitter tries to agree on a common key with the legitimate receivers while keeping it concealed from the eavesdropper. Simultaneously, the transmitter agrees on a private key with each of the legitimate receivers that needs to be kept secret from the other legitimate receiver and the eavesdropper. For this setup, we derive inner and outer bounds on the secret key capacity region. In the second setup, we assume that a backward public channel is available among the receivers and the transmitter. Each legitimate receiver wishes to share a private key with the transmitter. For this setup, an inner bound on the private key capacity region is found. Furthermore, the capacity region of the secret key in the state-dependent wiretap channel can be deduced from our inner and outer bounds.
\end{abstract}

% keywords
%\begin{keywords}
%Information theoretic security, 3-receiver broadcast channel, secret key sharing, channel model, secret key capacity.
%\end{keywords}

% \fi
%\IEEEpeerreviewmaketitle

\section{Introduction}{
In 1949, Shannon introduced a perfect secrecy condition in a system with an eavesdropper \cite{1}. The system is considered to be perfectly secure if $H(K) \geq H(M)$, where $H(K)$ and $H(M)$ are the entropies of the message and the key, respectively. In order to share keys in a network, common randomness is required to be distributed among legitimate users. Common randomness can be achieved through correlated sources or channel distribution, which categorizes secret key agreement into two different models; Source model and Channel model. The problem of secret key agreement in the source model was first studied by Ahlswede and Csiszar \cite{2}. They considered a network with two legitimate users and an eavesdropper. The legitimate users, which are connected together via an insecure noiseless channel, intend to agree on a secret key while observing correlated sources. In a noisy channel, common randomness  can be obtained using the channel distribution. This model is not beneficial for secret key sharing if the legitimate users do not have any advantages compared to the illegal users. Maurer solved this problem using a backward public channel in the wiretap model \cite{3}. In \cite{4}, Gohari and Anantharam studied the channel model for a multiple terminals network at which legitimate terminals and an eavesdropper are all connected to an interactive public channel. In addition, the terminals have access to a noisy M-receiver broadcast channel. The legitimate terminals agree on a common key using received messages from the M-receiver broadcast channel and the public channel. Instead of using the public channel, correlated sources can be utilized in the channel model. Khisti \emph{et al.} established this idea for key agreement over a wiretap channel where the transmitter and the legitimate receiver have access to correlated sources \cite{5}. Salimi and Skoglund developed the channel model with correlated sources in a generalized multiple access channel at which each of the transmitters intends to agree on an independent private key with the receiver \cite{6}.
In the channel model, assumption of the channel state information accessibility is more practical than existence of correlated sources. In such networks, the channel state information can be utilized to share secret keys. In \cite{7}-\cite{9}, the problem of secret key sharing is studied over the wiretap channel with non-causal and causal channel state information.

%\subsection*{Main Contributions and Organization}{
Consider a network with four types of users, trusted center, cluster node, end node and illegal user, which have different levels of accessibility. This network can be divided into sub-networks that each contains one cluster node and some end nodes. The trusted center broadcasts data signals to all users, while control signals are only transmitted to cluster nodes. The control signals that are sent to each cluster node must be concealed from the other cluster nodes. In addition, the illegal users try to eavesdrop both data and control signals. Motivated by the above scenario, we study the problem of secret key sharing in state-dependent 3-receiver broadcast channels where the Channel State Information (CSI) is non-causally available at the transmitter. We consider two cluster nodes, a trusted center and an illegal user. Without loss of generality end nodes are not considered. Two different models are discussed:
\begin{itemize}
\item \emph{Without Public Feedback}: The transmitter intends to agree on a common key with the legitimate receivers and a private key with the corresponding receivers.
\item \emph{With One Round Public Feedback}: An \emph{insecure} public channel is available among the receivers and the transmitter. Each legitimate receiver wishes to agree on a private key with the transmitter.
\end{itemize}

In the first model, an inner bound to the secret key capacity region is derived where the joint source channel coding is used. The coding scheme combines the aspects of both source and channel coding for secret key generation. Also, an outer bound to the secret key capacity region is obtained. In the second model, we establish an inner bound on the secret key capacity region where the double random binning is employed. Furthermore, it is shown that the capacity region of the secret key in the state-dependent wiretap channel can be deduced from our inner and outer bounds. 

The rest of the paper is organized as follows. In Section II, the system model is described. In Section III, our main results and the intuitions behind them are given. Finally, proof of theorems are illustrated in Section IV.

}
\section{Problem Definition}{
Throughout the paper, we denote a discrete random variable by an upper case letter (e.g., $X$) and its realization by the lower case letter (e.g., $x$). We denote the probability density function of $X$ over $\mathcal{X}$ by $p_{X}(x)$ and the conditional probability density function of $Y$ given $X$ by ${p_{Y|X}(y|x)}$. Finally, we use $X^n$ to indicate vector ${(X_{1},X_{2}, \ldots ,X_{n})}$.

In order to discuss the problem of secret key agreement over state-dependent 3-receiver broadcast channels, we consider a channel model consisting of a transmitter, two legitimate receivers and an eavesdropper. The transmitter communicates with the other three users over a 3-receiver broadcast channel. We assume that the channel is discrete memoryless with an input alphabet $\mathcal{X}$, output alphabet {$\mathcal{Y}_1 \times \mathcal{Y}_2 \times \mathcal{Z}$} depending on a parameter $S$ with values in a set $\mathcal{S}$.

In the described model, we consider two cases separately; A. no public feedback is allowed among the transmitter and the receivers. B. one round public feedback is allowed among the users. 

\subsection{The Model Without Public Feedback}{

In the model without public feedback, as Fig.~1 illustrates, the CSI is available at the transmitter. The transmitter, upon observing $s^n$, generates $k_0 \in {[1:2^{nR_0})}$ as a common key, and also, determines two independent keys $k_1 \in {[1:2^{nR_1})}$ and $k_2 \in {[1:2^{nR_2})}$ as private keys for sharing with the first and second legitimate receivers, respectively. After that, the transmitter determines the channel input $x_i$ as a (potentially random) function of the keys and $s^n$ for ${i \in [1:n]}$. Consequently, the outputs $y_1^n$, $y_2^n$ and $z^n$ are observed by the receivers. For $j=1,2$, the $j$th legitimate receivers estimates the keys $\hat{k}_0$ and $\hat{k}_j$ by means of its observation from the channel.
\begin{figure}
\includegraphics[width=3.1in,keepaspectratio]{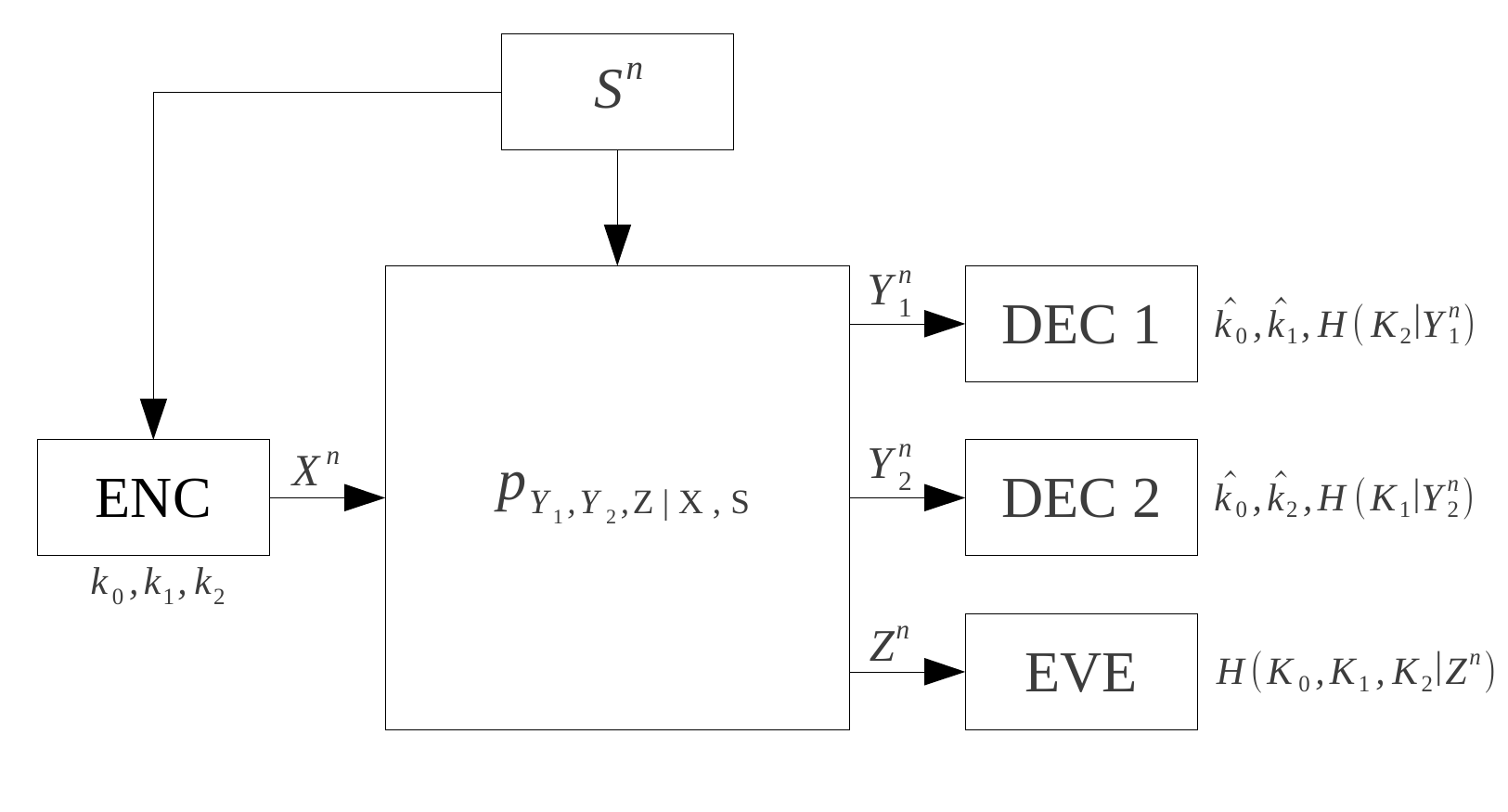}
\caption{{The state-dependent 3-receiver broadcast channel without public feedback}}
\end{figure}
\begin{definition}
A rate triple $(R_0,R_1,R_2)$ is said to be achievable if for every $\epsilon > 0$ and sufficiently large $n$, there exists a protocol such that\\
\vspace{-4mm}
\begin{align}
& \Pr \{ K_0 \neq \hat{K}_{0} \} < \epsilon, \label{1} \\
& \Pr \{ K_1 \neq \hat{K}_1 \} < \epsilon \ , \ \Pr \{ K_2 \neq \hat{K}_2 \} < \epsilon, \label{3} \\
&\frac{1}{n} I(K_0,K_1,K_2;Z^n) < \epsilon,  \label{6} \\
&\frac{1}{n} I(K_1;Y^n_2) < \epsilon, \label{4} \\
&\frac{1}{n} I(K_2;Y^n_1) < \epsilon, \label{5} \\
&\frac{1}{n} H(K_i) > {R_i}- \epsilon, \qquad \qquad \ \ \ \ \text{for} \ \ i=0,1,2. \label{8}
\end{align}
Equations (\ref{1}) and (\ref{3}) are the reliability conditions of the keys. Equation (\ref{6}) shows that the eavesdropper can not reconstruct the keys. Equations (\ref{4}) and (\ref{5}) mean that each legitimate receiver has efficiently no information about the other legitimate receiver's private key. Finally, the equation (\ref{8}) imposes uniformity condition.
\end{definition}
\begin{definition}
The secret key capacity region is the set of all achievable rate triples ${(R_0,R_1,R_2)}$.
\end{definition}

\subsection{The Model With One Round Public Feedback}{
This model has some differences with the model discussed in sub-section II.~A. In the current model, the legitimate receivers use the public feedback channel to transmit the required information to the transmitter for the key reconstruction. However, in the model without public channel, the transmitter transmits the required information via the 3-receiver broadcast channel to the legitimate receivers.

In the model with one round public feedback, as Fig.~2 illustrates, the CSI is also known non-causally at the transmitter and a public channel is available from the receivers to the transmitter. The transmitter, upon observing $s^n$, sends $x_i$ for {$i \in [1:n]$} over the channel. For $j=1,2$, the $j$th legitimate receiver transmits $\psi_j$ as (potentially random) function of its observation $y_j^n$, using the public feedback channel. Then, the $j$th legitimate receiver generates private key {$k_j \in {[1:2^{nR_j})}$} as (potentially random) function of $y_j^n$, $\psi_1$ and $\psi_2$. The transmitter determines $\hat{k}_1$ and $\hat{k}_2$ as the estimation of the keys $k_1$ and $k_2$ by means of the received messages from the public feedback channel and {$(x^n,s^n)$}.

\begin{definition}
A rate pair $(R_1,R_2)$ is said to be achievable if for every $\epsilon > 0$ and sufficiently large $n$, there exists a protocol such that
\begin{align}
& \Pr \{ K_1 \neq \hat{K}_1 \} < \epsilon \ , \ \Pr \{ K_2 \neq \hat{K}_2 \} < \epsilon, \label{10} \\
&\frac{1}{n} I(K_1;Y^n_2,\psi_1,\psi_2) < \epsilon, \label{11} \\
&\frac{1}{n} I(K_2;Y^n_1,\psi_1,\psi_2) < \epsilon, \label{12}  \\
&\frac{1}{n} I(K_1,K_2;Z^n,\psi_1,\psi_2) < \epsilon, \label{13} \\
&\frac{1}{n} H(K_i) > {R_i}- \epsilon, \qquad \qquad \ \ \ \ \text{for} \ \ i=1,2.  \label{15}
\end{align}
\end{definition}
Equation (\ref{10}) is the reliability conditions of the private keys. Equations (\ref{11}) and (\ref{12}) mean that each legitimate receiver has efficiently no information about the other legitimate receiver's private key. Equation \eqref{13} illustrates that the eavesdropper can not reconstruct the private keys. Finally, the equation (\ref{15}) imposes uniformity condition.

\begin{definition}
The private key capacity region is the set of all achievable rate pairs $(R_1,R_2)$.
\end{definition}
}
\section{Main Results}{
In this section, we state the main results about the described models. We discuss the cases of without public feedback in sub-section III.~A and with one round public feedback in sub-section III.~B.
\subsection{The Model Without Public Feedback}{
For the model without public feedback, we establish the following inner and outer bounds on the secret key capacity region.
\begin{theorem}[Inner Bound]
The rate triple $(R_0,R_1,R_2)$ is achievable for the model without public feedback if: 
\begin{align}
R_0 &\leq [\min \{I(U_0;Y_1) , I(U_0;Y_2) \}-I(U_0;Z)]^{+},  \nonumber \\
R_1 &\leq [I(U_1;Y_1|U_0)-I(U_1;Y_2,U_2|U_0)]^+, \nonumber \\
R_2 &\leq [I(U_2;Y_2|U_0)-I(U_2;Y_1,U_1|U_0)]^+, \nonumber \\
R_0+R_1 &\leq [ \min \{I(U_0;Y_1) , I(U_0;Y_2) \}, \nonumber \\
&+I(U_1,Y_1|U_0) - I(U_0,U_1;Z)]^+, \nonumber \\
R_0+R_2 &\leq [\min \{I(U_0;Y_1) , I(U_0;Y_2) \}, \nonumber \\
&+I(U_2,Y_2|U_0) - I(U_0,U_2;Z)]^+, \nonumber \\
R_0+R_1+R_2 &\leq [\min \{I(U_0;Y_1) , I(U_0;Y_2) \}, \nonumber \\ 
&+I(U_1,Y_1|U_0) +I(U_2,Y_2|U_0), \nonumber \\ 
&- I(U_0,U_1,U_2;Z)-I(U_1;U_2|U_0)]^+, \nonumber
\end{align} 
%&R_0+R_1+R_2 \leq \min \{I(U_0;Y_1),I(U_0;Y_2) \} + I(U_1;Y_1|U_0)+I(U_2;Y_2|U_0)-I(U_0,U_1,U_2;Z)
subject to the constraints:
\begin{align}
I(U_0;Y_1) &\geq I(U_0;S), \nonumber \\ 
I(U_0;Y_2) &\geq I(U_0;S), \nonumber \\
I(U_1;Y_1|U_0) &\geq I(U_1;S|U_0), \nonumber \\
I(U_2;Y_2|U_0) &\geq I(U_2;S|U_0),
\end{align} 
\vspace{-3mm}
for some input distributions:
\begin{align}
&p_{S,U_0,U_1,U_2,X,Y_1,Y_2,Z}(s,u_0,u_1,u_2,x,y_1,y_2,z)= \nonumber \\
&p_{S}(s)p_{U_0|S}(u_0|s)p_{U_1|S,U_0}(u_1|s,u_0)p_{U_2|S,U_0}(u_2|s,u_0) \nonumber \\
&p_{X|S,U_0,U_1,U_2}(x|s,u_0,u_1,u_2)p_{Y_1,Y_2,Z|X,S}(y_1,y_2,z|x,s), \nonumber
\end{align} 
where the function $[x]^+$ equals to $x$ if $x \geq 0$ and $0$ if ${x<0}$.
\emph{Proof}: See Section IV.~A. 
\end{theorem}
\begin{figure}
\includegraphics[width=3.2in,keepaspectratio]{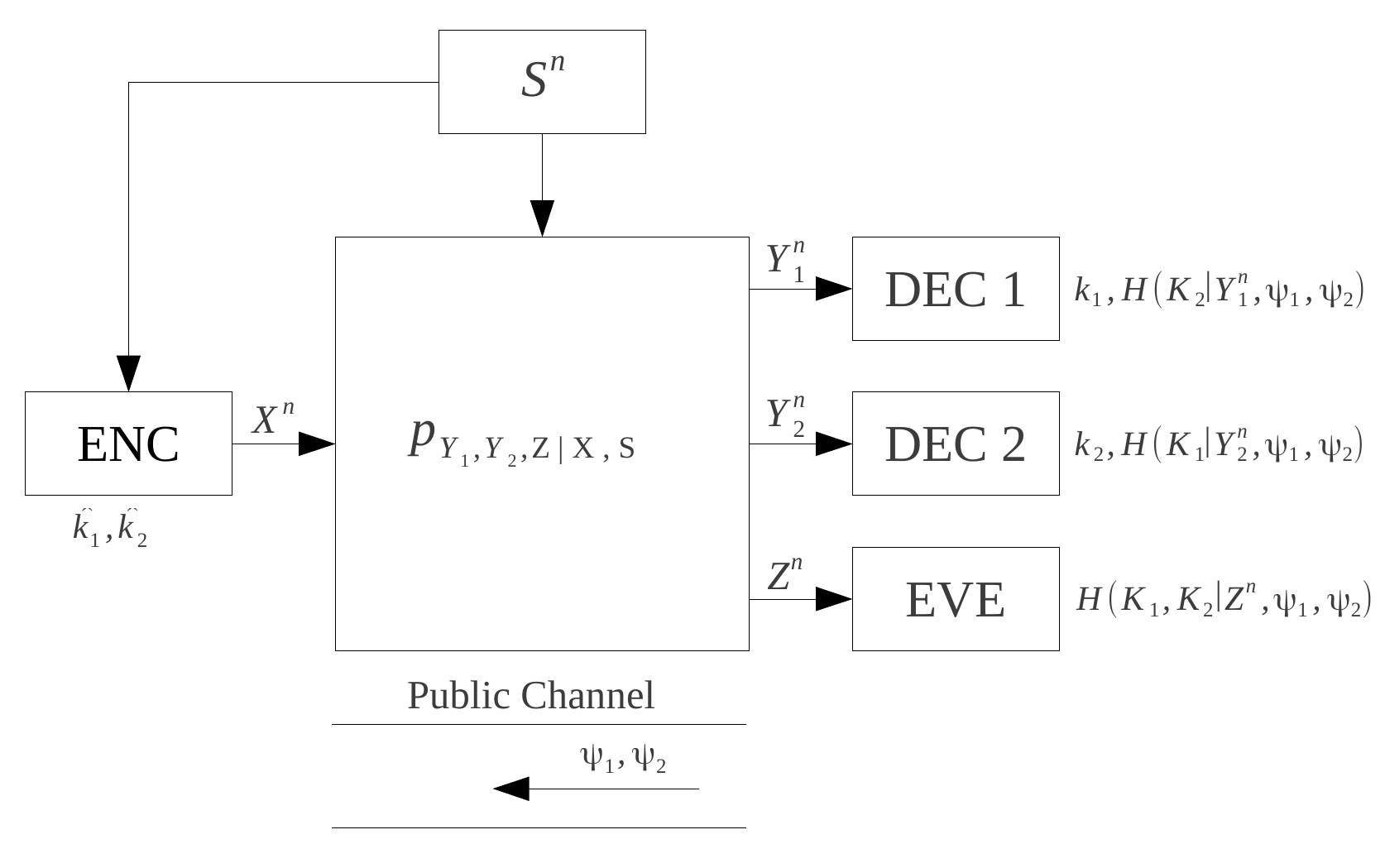}
%\vspace{-0.4 cm}
\caption{The state-dependent 3-receiver broadcast channel with presence of one round public feedback}
\end{figure}

For the achievability, we use a scheme which utilizes the joint source channel coding for two discrete memoryless sources over the broadcast channel and also, random binning to satisfy secrecy constrains.
\begin{theorem}[Outer Bound]
For the model without public feedback, any rate triple $(R_0,R_1,R_2)$ must satisfy
\begin{align}
&R_0 \leq \min \{I(X,S;Y_1|Z),I(X,S;Y_2|Z) \}, \nonumber \\
&R_1 \leq \min \{I(X,S;Y_1|Y_2),I(X,S;Y_1|Z) \}, \nonumber \\
&R_2 \leq \min \{I(X,S;Y_2|Y_1),I(X,S;Y_2|Z) \},
\end{align} 
%&R_0+R_1+R_2 \leq I(X,S_1,S_2;Y_1,Y_2|Z)
for some input distributions $p_{S,X,Y_1,Y_2,Z}$.\\
\emph{Proof}: See Section IV.~B.
\end{theorem}
\emph{Remark~1:} By setting ${U_1=U_2=\emptyset}$ in Theorem 1 and $Y_1=Y_2=Y$ in Theorem 1 and Theorem 2, the region reduces to the region of the secret key in the wiretap channel without public feedback \cite{7}.
%\vspace{-2mm}
\subsection{The Model With One Round Public Feedback}{
For the model with one round public feedback, we establish the following inner bound on the private key capacity region.

\begin{theorem}[Inner Bound]
The rate pair $(R_1,R_2)$ is achievable for the model with one round public feedback if:
\begin{align}
R_1 \leq [&\min \{I(X,S;V_1)-I(V_1;Y_2),\nonumber \\
&I(X,S;V_1)-I(V_1;Z) \}]^+  \nonumber \\
R_2 \leq [&\min \{I(X,S;V_2)-I(V_2;Y_1),\nonumber \\
&I(X,S;V_2)-I(V_2;Z) \}]^+ \nonumber \\
R_1+R_2 \leq I&(X,S;V_1,V_2)-I(V_1,V_2;Z), \label{26}
\end{align} 
%& R_1+R_2 \leq I(X,S;Y_1)+I(X,S;Y_2)-I(Y_1,Y_2;Z)
%\vspace{-2mm}
for some input distributions:
\begin{align}
&p_{S,X,V_1,V_2,Y_1,Y_2,Z}(s,x,v_1,v_2,y_1,y_2,z)=p_{S}(s)p_{X|S}(x|s) \nonumber \\
&p_{Y_1,Y_2,Z|X,S}(y_1,y_2,z|x,s)p_{V_1|Y_1}(v_1|y_1)p_{V_2|Y_2}(v_2|y_2).\nonumber
\end{align} 
\end{theorem}
\emph{Proof}: See Section IV.~C.

The main idea of achieving the inner bound comes from the Slepian and Wolf coding scheme for distributed lossless source coding problems \cite{10}. In addition, we use the double random binning to satisfy the secrecy constrains.

\emph{Remark~2:} By setting ${Y_1=V_1= \emptyset}$ and $Y_2=V_2=Y$ (or $Y_2=V_2=\emptyset$ and $Y_1=V_1=Y$) in Theorem 3, the inner bound reduces to the inner bound on the secret key in the wiretap channel with one round public feedback \cite{8}.
}
%\vspace{-1.5mm}
\section{Proofs}{
In this section, we illustrate the proof of theorems. In order to prove Theorem 1, the hybrid joint source channel coding and random binning are used. The innovation behind the proof of Theorem 3 comes from the Slepian and Wolf coding and the double random binning.
%\vspace{-1.75mm}
\subsection{Proof of Theorem 1}{
Fix conditional probability density function $p_{U_0,U_1,U_2|S}$.

\emph{Codebook Generation}: Generate $2^{n\tilde{R}_0}$ sequences $U^n_0(m_{u_0})$ {$m_{u_0} \in {[1:2^{n\tr_0})}$}, each according to $\prod_{i=1}^n p_{U_0}(u_{0i})$.~Then, randomly and independently partition them into $2^{nR_0}$ bins. For each $m_{u_0}$, we generate $2^{n\tr_1}$ sequences $U^n_1(m_{u_1})$, $m_{u_1} \in {[1:2^{n\tr_1})}$ each according to $\prod_{i=1}^n p_{U_1|U_0}(u_{1i}|u_{0i})$, and partition them into $2^{nR_1}$ bins. Similarly, for each $m_{u_0}$, we generate $2^{n\tr_2}$ sequences $U^n_2(m_{u_2}), \ \ m_{u_2} \in {[1:2^{n\tr_2})}$ each according to $\prod_{i=1}^n p_{U_2|U_0}(u_{2i}|u_{0i})$. At the end, randomly partition the sequences $U^n_2$ into $2^{nR_2}$ bins.
%\vspace{-3mm}

\emph{Encoding}: For each sequence $s^n$, the encoder chooses a triple $(m_{u_0},m_{u_1},m_{u_2})$ such that $(s^n,u^n_0(m_{u_0}),u^n_1(m_{u_1}),$ $u^n_2(m_{u_2})) \in \mt^{(n)}_\epsilon(S,U_0,U_1,U_2)$. Then, the transmitter sends $x_i$ according to $p_{X|{U_0},{U_1},{U_2},S}(x_i|u_{0i},u_{1i},u_{2i},s_i)$ for ${i \in [1:n]}$ over the 3-receiver broadcast channel. By the covering lemma \cite{11}, this can be done with an arbitrarily small probability of error if:
\begin{align}
&\tr_0 \geq I(U_0;S), \nonumber \\
&\tr_1 \geq I(U_1;S|U_0), \nonumber \\
&\tr_2 \geq I(U_2;S|U_0).
\end{align} 

\emph{Decoding}: After receiving $y^n_1$, the first legitimate receiver finds a sequence pair $(u^n_0(\hm_{u_0}),u^n_1(\hm_{u_1}))$, such that $(u^n_0(\hm_{u_0}),u^n_1(\hm_{u_1}),y^n_1)$ $\in$ $\mt^{(n)}_{\epsilon}(U_0,U_1,Y_1)$. Similarly, upon observing $y^n_2$ the second legitimate receiver finds a sequence pair $(u^n_0(\hm_{u_0}),u^n_2(\hm_{u_2}))$, such that $(u^n_0(\hm_{u_0}),$ $u^n_2(\hm_{u_2}),y^n_2) \in \mt^{(n)}_{\epsilon}(U_0,U_2,Y_2)$. By the packing lemma \cite{11}, the probability of error tends to zero as $n \to \infty$ if:
\begin{align}
&\tr_0 \leq \min \{I(U_0;Y_1),I(U_0;Y_2)\}, \nonumber \\
&\tr_1 \leq I(U_1;Y_1|U_0), \nonumber \\
&\tr_2 \leq I(U_2;Y_2|U_0).
\end{align} 

\emph{Secret Key Generation}: After decoding, the transmitter and the legitimate receivers agree on the bin index of $u^n_0$ to be the public key $k_0$. The transmitter and the first legitimate receiver agree on the bin index of $u^n_1$ as the private key, likewise, the second legitimate receiver agrees on the bin index of $u^n_2$ as the private-key $k_2$ with the transmitter.

\emph{Analysis of Secrecy}: In order to check the secrecy condition \eqref{6}, we have:
{\small{\begin{align}
I(K_0,K_1,K_2;Z^n|\mc)&=I(K_0;Z^n|\mc)+I(K_1;Z^n|K_0,\mc) \nonumber \\
&+I(K_2;Z^n|K_0,K_1,\mc), \label{20}
\end{align}}}to satisfy the secrecy condition all the above terms must tend to zero as $n \to \infty$, for the first term we have:
{\small{\begin{align}
&I(K_0;Z^n|\mathcal{C}) = H(K_0|\mathcal{C})-H(K_0|Z^n,\mathcal{C})= H(K_0|\mathcal{C}) \qquad \qquad \qquad \qquad \nonumber \\
&-H(K_0,U^n_0|Z^n,\mathcal{C})+H(U^n_0|K_0,Z^n,\mathcal{C})= H(K_0|\mathcal{C}) \nonumber
\end{align}
\begin{align}
&-H(U^n_0|Z^n,\mathcal{C})-H(K_0|U^n_0,Z^n,\mathcal{C})+H(U^n_0|K_0,Z^n,\mathcal{C}) \qquad \qquad \qquad \nonumber \\
&\stackrel{(a)}{=} H(K_0|\mathcal{C})-H(U^n_0|Z^n,\mathcal{C})+H(U^n_0|K_0,Z^n,\mathcal{C}) \nonumber \\
&\stackrel{(b)}{\leq}nR_0-H(U^n_0|Z^n,\mathcal{C})+H(U^n_0|K_0,Z^n,\mathcal{C})\stackrel{(c)}{\leq} nR_0 \nonumber \\
&-n\tilde{R}_0+nI(U_0;Z)+H(U^n_0|K_0,Z^n,\mathcal{C}) \nonumber \\
&\stackrel{(d)}{\leq} nR_0-n\tilde{R}_0+nI(U_0;Z)+n\tilde{R}_0-nR_0 \nonumber \\
&-nI(U_0;Z)+n\epsilon = n\epsilon, \nonumber \end{align}}}as $K_0$ is the bin index of $U^n_0$, the equality $H(K_0|U^n_0,Z^n, \mathcal{C})\\=0$ holds which implies $(a)$. $(b)$ comes from the fact that $H(K_0|\mathcal{C}) \leq H(K_0) \leq nR_0$. $(c)$ can be deduced from inequality $H(U^n_0|Z^n,\mathcal{C}) \geq n\tilde{R}_0-nI(U_0;Z)$.
{\small{\begin{align}
&H(U_0^n|Z^n,\mathcal{C})= H(U_0^n,Z^n|\mathcal{C})-H(Z^n|\mathcal{C})=H(U_0^n|\mathcal{C}) \nonumber \\
&+H(Z^n|U_0^n,\mc)-H(Z^n|\mathcal{C})=n \tilde{R}_0+H(Z^n|U_0^n,\mc)-H(Z^n|\mathcal{C}) \nonumber \\
&=n \tr_0 - I(Z^n;U_0^n|\mc) \geq n \tr_0 - I(Z^n;U_0^n,\mc) =  n \tilde{R}_0 \nonumber \\
&-I(Z^n;U_0^n) = n \tilde{R}_0 - nI(Z;U_0), \nonumber
\end{align}}}the inequality $H(U^n_0|K_0,Z^n,\mathcal{C}) \leq n\tilde{R}_0-nR_0-nI(U_0;Z)+n\epsilon$ is established if $\tr_0-R_0 \geq I(U_0;Z)+\epsilon$ which implies $(d)$. The proof is similar to Appendix~22C in \cite{11}. For the second term of \eqref{20},
{\small{\begin{align}
&I(K_1;Z^n|K_0,\mc)=H(K_1|K_0,\mc)-H(K_1|K_0,Z^n,\mc) \stackrel{(a)}{\leq} nR_1  \nonumber \\ 
&-H(K_1,U_1^n,U_0^n|K_0,Z^n,\mc)+H(U_1^n,U_0^n|K_0,K_1,Z^n,\mc) \nonumber \\
&\stackrel{(b)}{=} nR_1 - H(U_1^n,U_0^n|K_0,Z^n,\mc)+H(U_1^n,U_0^n|K_0,K_1,Z^n,\mc) \nonumber \\
&=nR_1 - H(U_0^n|K_0,Z^n,\mc)- H(U_1^n|K_0,U_0^n,Z^n,\mc)  \nonumber \\ 
&+H(U_1^n,U_0^n|K_0,K_1,Z^n,\mc)\stackrel{(c)}{\leq} nR_1 - n(\tr_0-R_0)+nI(U_0;Z) \nonumber \\
&-n\tr_1 +nI(U_1;Z|U_0)+H(U_1^n,U_0^n|K_0,K_1,Z^n,\mc)  \nonumber \\ 
&\stackrel{(d)}{\leq} nR_1 - n(\tr_0-R_0) + nI(U_0;Z)-n\tr_1+nI(U_1;Z|U_0) \nonumber \\
&+n(\tr_0-R_0)+n(\tr_1-R_1)-nI(U_0,U_1;Z)+n\epsilon = n \epsilon, \nonumber
\end{align}}}where $(a)$ comes from the fact that $H(K_1|K_0,\mathcal{C}) \leq H(K_1) \leq nR_1$. As $K_1$ is the bin index of $U^n_1$, the equality $H(K_1|K_0,$ $U_1^n,U^n_0,Z^n,\mathcal{C})=0$ holds, which implies $(b)$. In order to prove $(c)$, we have:
{\small{\begin{align}
&H(U_0^n|K_0,Z^n,\mathcal{C})= H(U_0^n,Z^n|K_0,\mathcal{C})-H(Z^n|K_0,\mathcal{C}) \nonumber \\
&=H(U_0^n|K_0,\mathcal{C})+H(Z^n|K_0,U_0^n,\mathcal{C})-H(Z^n|K_0,\mathcal{C}) \nonumber \\ 
&=n (\tilde{R}_0-R_0)-I(Z^n;U_0^n|K_0,\mc)\geq n(\tilde{R}_0-R_0) \nonumber \\
&-I(Z^n;U_0^n,K_0,\mc)=n(\tilde{R}_0-R_0)-I(Z^n;U_0^n) \nonumber \\
&-I(Z^n; K_0,\mc|U_0^n)= n(\tilde{R}_0-R_0) - I(Z^n;U_0^n)= n(\tilde{R}_0-R_0) \qquad \nonumber \\
&-nI(Z;U_0), \nonumber
\end{align}}}and
{\small{\begin{align}
&H(U_1^n|K_0,U_0^n,Z^n,\mathcal{C})= H(U_1^n,Z^n|K_0,U_0^n,\mathcal{C}) \nonumber \\
&-H(Z^n|K_0,U_0^n,\mathcal{C})=H(U_1^n|K_0,U_0^n,\mathcal{C}) \nonumber \\
&+H(Z^n|K_0,U_0^n,U_1^n,\mathcal{C})-H(Z^n|K_0,U_0^n,\mathcal{C}) \nonumber \\ 
&=n \tilde{R}_1-I(Z^n;U_1^n|K_0,U_0^n, \mc) \geq n\tilde{R}_1-I(Z^n;U_1^n,K_0,\mc|U_0^n)  \nonumber \qquad \\
&= n \tilde{R}_1 - I(Z^n;U_1^n|U_0^n)- I(Z^n; K_0,\mc|U_1^n,U_0^n) \nonumber \\ 
&= n \tilde{R}_1 - I(Z^n;U_1^n|U_0^n)= n \tilde{R}_1 - nI(Z;U_1|U_0), \nonumber
\end{align}}}the inequality $H(U_1^n,U_0^n|K_0,K_1,Z^n,\mc) \leq n(\tr_0-R_0)+n(\tr_1-R_1)-nI(U_0,U_1;Z)+n\epsilon$ is established if $(\tr_0-R_0)+(\tr_1-R_1) \geq I(U_0,U_1;Z)+\epsilon$ which implies $(d)$, the proof is similar to Appendix~22C in \cite{11}. For the third term of \eqref{20}, we have:
{\small{\begin{align}
&I(K_2;Z^n|K_0,K_1,\mc)=H(K_2|K_0,K_1,\mc) \nonumber \\ 
&-H(K_2|K_0,K_1,Z^n,\mc) \nonumber \\
&\stackrel{(a)}{\leq} nR_2 -H(K_2,U_2^n,U_1^n,U_0^n|K_0,K_1,Z^n,\mc) \nonumber \\
&+H(U_2^n,U_1^n,U_0^n|K_0,K_1,K_2,Z^n,\mc) \nonumber \\ 
&\stackrel{(b)}{=} nR_2 - H(U_2^n,U_1^n,U_0^n|K_0,K_1,Z^n,\mc)\nonumber \\
&+H(U_2^n,U_1^n,U_0^n|K_0,K_1,K_2,Z^n,\mc) \nonumber \\
&= nR_2 - H(U_1^n,U_0^n|K_0,K_1,Z^n,\mc) \nonumber \\
&- H(U_2^n|K_0,K_1,U_1^n,U_0^n,Z^n,\mc) \nonumber \\
& +H(U_2^n,U_1^n,U_0^n|K_0,K_1,K_2,Z^n,\mc) \stackrel{(c)}{\leq} nR_2-n(\tr_0-R_0) \qquad \nonumber \\
&-n(\tr_1-R_1)+nI(U_1,U_0;Z)-n\tr_2+nI(U_2;Z,U_1|U_0) \nonumber \\
&+H(U_2^n,U_1^n,U_0^n|K_0,K_1,K_2,Z^n,\mc) \stackrel{(d)}{\leq} nR_2- n(\tr_0-R_0) \nonumber \\
&-n(\tr_1-R_1)+nI(U_1,U_0;Z)-n\tr_2+nI(U_2;Z,U_1|U_0) \nonumber \\
&+n(\tr_0-R_0)+n(\tr_1-R_1)+n(\tr_2-R_2)-nI(U_1,U_0;Z) \nonumber \\
&-nI(U_2;Z,U_1|U_0)+n\epsilon = n \epsilon, \nonumber
\end{align}}}where $(a)$ comes from the fact that $H(K_2|K_0,K_1,\mathcal{C}) \leq H(K_2) \leq nR_2$. As $K_2$ is the bin index of $U^n_2$, the equality $H(K_2|K_0,K_1,U_2^n,U_1^n,U^n_0,Z^n,\mathcal{C})=0$ holds which implies $(b)$. In order to prove $(c)$, we have: {\small{\begin{align}
&H(U_1^n,U_0^n|K_0,K_1,Z^n,\mathcal{C})= H(U_1^n,U_0^n,Z^n|K_0,K_1,\mathcal{C}) \nonumber \\
&-H(Z^n|K_0,K_1,\mathcal{C})=H(U_1^n,U_0^n|K_0,K_1,\mathcal{C}) \nonumber \\
&+H(Z^n|K_0,K_1,U_1^n,U_0^n,\mathcal{C})-H(Z^n|K_0,K_1,\mathcal{C}) \nonumber \\
&=n (\tilde{R}_0-R_0)+n (\tilde{R}_1-R_1) -I(Z^n;U_1^n,U_0^n|K_0,K_1, \mc) \nonumber \qquad \\
&\geq n(\tilde{R}_0-R_0)+n (\tilde{R}_1-R_1) -I(Z^n;U_1^n,U_0^n,K_0,K_1,\mc) \nonumber \\
&= n(\tilde{R}_0-R_0)+n (\tilde{R}_1-R_1) - I(Z^n;U_1^n,U_0^n) \nonumber \\
&-I(Z^n; K_0,K_1,\mc|U_1^n,U_0^n)= n(\tilde{R}_0-R_0)+n(\tilde{R}_1-R_1) \nonumber \\
&-I(Z^n;U_1^n,U_0^n) = n(\tilde{R}_0-R_0)+n (\tilde{R}_1-R_1) - nI(Z;U_1,U_0), \nonumber
\end{align}}}and
{\small{\begin{align}
&H(U_2^n|K_0,K_1,U_1^n,U_0^n,Z^n,\mathcal{C})= H(U_2^n,U_1^n,Z^n,K_1|K_0,U_0^n,\mathcal{C}) \nonumber \\
&-H(U_1^n,Z^n,K_1|K_0,U_0^n,\mathcal{C})=H(U_2^n|K_0,U_0^n,\mathcal{C}) \nonumber \\
&+H(U_1^n,Z^n,K_1|K_0,U_0^n,U_2^n,\mathcal{C})-H(U_1^n,Z^n,K_1|K_0,U_0^n,\mathcal{C}) \nonumber \\
&=n \tilde{R}_2-I(U_1^n,Z^n,K_1;U_2^n|K_0,U_0^n, \mc)=n \tilde{R}_2 \nonumber \\
&-I(U_1^n,Z^n;U_2^n|K_0,U_0^n, \mc)-I(K_1;U_2^n|K_0,U_1^n,U_0^n,Z^n, \mc) \nonumber \\
&\geq n\tilde{R}_2 -I(U_1^n,Z^n;U_2^n,K_0,\mc|U_0^n) = n \tilde{R}_2-I(U_1^n,Z^n;U_2^n|U_0^n) \nonumber \\
&- I(U_1^n,Z^n; K_0,\mc|U_2^n,U_0^n)= n \tilde{R}_2 - I(U_1^n,Z^n;U_2^n|U_0^n) \nonumber \\ 
&\geq n \tilde{R}_2-nI(U_1,Z;U_2|U_0), \qquad \qquad \qquad \qquad \qquad \qquad \qquad \qquad \qquad \nonumber
\end{align}}}the inequality $H(U_2^n,U_1^n,U_0^n|K_0,K_1,K_2,Z^n,\mc) \leq n(\tr_0-R_0)+n(\tr_1-R_1)+n(\tr_2-R_2)-nI(U_0,U_1;Z)-nI(U_2;Z,U_1|U_0)+n\epsilon $ is established if $(\tr_0-R_0)+(\tr_1-R_1)+(\tr_2-R_2) \geq I(U_0,U_1;Z)+I(U_2;Z,U_1|U_0)+\epsilon$ which implies $(d)$. The proof is similar to Appendix~22C in \cite{11}. By checking \eqref{4} and \eqref{5} in similar way, the following conditions are obtained:
%\vspace{-2mm}
\begin{align}
&\tr_1-R_1 \geq I(U_1;Y_2,U_2|U_0)+ \epsilon, \nonumber \\
&\tr_2-R_2 \geq I(U_2;Y_1,U_1|U_0)+ \epsilon. \nonumber
\end{align}
}
\vspace{-10mm}
\subsection{Proof of Theorem 2}{
In our described model, each legitimate receiver must be able to estimate the common key $k_0$ correctly and according to the Fano's inequality ${\frac{1}{n}H(K_0|Y^n_1) \leq \epsilon}$ and ${\frac{1}{n}H(K_0|Y^n_2) \leq \epsilon}$, and also, the secrecy condition ${\frac{1}{n}I(K_0;Z^n) \leq \epsilon}$ must be satisfied. We find an outer bound on $R_0$. (outer bounds on $R_1 \ \text{and} \ R_2$ can be obtained similarly)
{\small{ \begin{align}
&nR_0=H(K_0)=I(K_0;Y^n_1)+H(K_0|Y^n_1)\stackrel{(a)}{\leq} I(K_0;Y^n_1) + n {\epsilon}  \nonumber \\
& \leq I(K_0;Y^n_1,Z^n)+n {\epsilon}=I(K_0;Z^n)+I(K_0;Y^n_1|Z^n)+n{\epsilon} \nonumber \\
& \stackrel{(b)}{\leq} I(K_0;Y^n_1|Z^n) + 2n {\epsilon}\leq I(K_0,X^n,S^n;Y^n_1|Z^n)+2n{\epsilon} \nonumber  \\
&=I(X^n,S^n;Y^n_1|Z^n)+I(K_0;Y^n_1|Z^n,X^n,S^n)+2n{\epsilon} \nonumber \\
&\stackrel{(c)}{\leq}I(X^n,S^n;Y^n_1|Z^n)+2n{\epsilon} \leq \sum_{i=1}^{n}I(X_i,S_i;Y_{1i}|Z_i)+2n{\epsilon} \nonumber \\
& \stackrel{(d)}{=}nI(X_Q,S_Q;Y_{1Q}|Z_Q,Q)+2n{\epsilon} \leq nI(X_Q,S_Q,Q;Y_{1Q}|Z_Q) \qquad \nonumber \\
& +2n{\epsilon} = nI(X_Q,S_Q;Y_{1Q}|Z_Q)+nI(Q;Y_{1Q}|X_Q,S_Q,Z_Q) \nonumber \\
&+2n{\epsilon}\stackrel{(e)}{=} nI(X_Q,S_Q;Y_{1Q}|Z_Q)+2n{\epsilon},  \nonumber
\end{align} }}where $(a)$ and $(b)$ comes from the Fano's inequality and the secrecy condition, respectively. $(c)$ can be deduced from the Markov chain $K_0$ $\rightarrow$ $(X^n,S^n)$ $\rightarrow$ $(Y^n_1,Y^n_2,Z^n)$ and $(d)$ can be  obtained by defining $Q$ as a uniform random variable over ${[1:n]}$. The Markov chain $Q$ $\rightarrow$ $(S_Q,X_Q)$ $\rightarrow$ $Y_{1Q}$ implies $(e)$. Also in a similar way, we have $R_0 \leq I(X,S;Y_2|Z)$. Finally, $R_0 \leq \min \{I(X,S;Y_1|Z),I(X,S;Y_2|Z) \}$.
}
\subsection{Proof of Theorem 3}{
Fix probability distributions $p_{V_1}$ and $p_{V_2}$.

\emph{Codebook Generation}: Consider the set of all typical sequences $V^n_1$ and $V^n_2$ with probability distribution $\prod_{i=1}^{n}p_{V_1}(v_{1i}) $ and $\prod_{i=1}^{n}p_{V_2}(v_{2i})$. The number of sequences $v^n_1$ and $v^n_2$ are $2^{n(H(V_1)+\epsilon_1)}$ and $2^{n(H(V_2)+\epsilon_2)}$, respectively, where $\epsilon_1 , \epsilon_2 >0$ can be chosen arbitrarily small. For $j=1,2$, partition sequences $v^n_j$ into $2^{nR'_j}$ bins and each bin into $2^{nR_j}$ sub-bins, using double random binning. Therefore, in average, there are $2^{n(H(V_j)-R'_j+\epsilon_j)}$ and $2^{n(H(V_j)-R'_j-R_j+\epsilon_j)}$ sequences $v_j^n$ in each bin and sub-bin, respectively.

\emph{Encoding}: Upon observing $s^n$, the transmitter randomly and independently generates a codeword $x^n$ according to the conditional distribution {\small {${\prod_{i=1}^n p_{X|S}(x_i|s_i)}$,}} and transmits $x_i$ for ${i \in [1:n]}$ over the 3-receiver broadcast channel with the channel probability distribution ${p_{Y_1,Y_2,Z|X,S}}$.

\emph{Use of The Public Channel}: On the basis of channel output sequence $y^n_1$ the first legitimate receiver finds $v^n_1$ such that $(v^n_1, y^n_1) \in \mt^{(n)}(V_1,Y_1)$ and sends the bin index of $v^n_1$ via the public channel. Similarly, the second legitimate receiver finds the bin index of $v^n_2$ such that $(v^n_2, y^n_2) \in \mt^{(n)}(V_2,Y_2)$ and transmits the bin index of $v^n_2$ over the public channel.

\emph{Secret Key Generation}: The transmitter upon receiving the bin indices of $v^n_1$ and $v^n_2$ from the public channel and by the knowledge of $x^n$ and $s^n$ can recover $v^n_1$ and $v^n_2$. Then both transmitter and the first legitimate receiver agree on the sub-bin index of $v^n_1$ as the private key. Similarly, the second legitimate receiver and the transmitter agree on the sub-bin index of $v^n_2$ as their private key.

\emph{Analysis of The Probability of Error}: From the Slepian and Wolf theorem, the probability of error tends to zero as $n \to \infty$ if:
\vspace{-2mm}
\begin{align}
&R'_1 \geq H(V_1|X,S), \  R'_2 \geq H(V_2|X,S), \nonumber \\
&R'_1+R'_2 \geq H(V_1,V_2|X,S). \label{25}
\end{align} 

\emph{Analysis of Secrecy}: We will prove that the following constraints are sufficient to satisfy the secrecy condition \eqref{13}.
\begin{align}
&R_1+R'_1 \leq H(V_1|Z), \ R_2+R'_2 \leq H(V_2|Z), \nonumber  \\
&R'_1+R'_2+R_1+R_2 \leq H(V_1,V_2|Z). \label{23}
\end{align}
In order to prove \eqref{23}, we have:
{\small{\begin{align}
I(K_1,K_2;Z^n,\psi_1,\psi_2|\mc)&= I(K_1;Z^n,\psi_1|\mc)+I(K_1;\psi_2|Z^n,\psi_1,\mc) \nonumber \\
&+I(K_2;Z^n,\psi_1,\psi_2|K_1,\mc). \label{21}
\end{align}}}For the first term we have:
{\small{\begin{align}
&I(K_1;Z^n,\psi_1|\mathcal{C})=I(K_1,V^n_1;Z^n,\psi_1|\mathcal{C})-I(V^n_1;Z^n,\psi_1|K_1,\mathcal{C}) \qquad \nonumber \\
&\stackrel{(a)}{=} I(V^n_1;Z^n,\psi_1|\mathcal{C})-H(V^n_1|K_1,\mathcal{C})+H(V^n_1|Z^n,\psi_1,K_1,\mathcal{C}) \nonumber \\
&=H(V^n_1|\mc)-H(V^n_1|Z^n,\psi_1,\mathcal{C})-H(V^n_1|K_1,\mathcal{C}) \nonumber \\
&+H(V^n_1|Z^n,\psi_1,K_1,\mathcal{C}), \nonumber 
\end{align}}}as $K_2$ is the sub-bin index of $V_1^n$ the equality $I(K_1;Z^n,$ $\psi_1|V_1^n,\mc)=0$ holds, which implies $(a)$. Now 
{\small{\begin{align}
&H(V^n_1|Z^n,\psi_1,\mathcal{C})=H(V^n_1|Z^n)+H(\psi_1|Z^n,V^n_1,\mathcal{C}) \nonumber \qquad \qquad \qquad \qquad \\
&-H(\psi_1|Z^n,\mathcal{C})\stackrel{(b)}{=}H(V^n_1|Z^n)-H(\psi_1|Z^n,\mathcal{C}) \nonumber \\ 
&\geq H(V^n_1|Z^n)-nR'_1 \nonumber=n(H(V_1|Z)-R'_1), \nonumber
\end{align}}}where $(b)$ comes from the fact that $\psi_1$ is the bin index of $V^n_1$,
{\small{\begin{align}
&H(V^n_1|K_1,\mathcal{C})=H(V^n_1)+H(K_1|V^n_1,\mathcal{C})-H(K_1|\mathcal{C})\stackrel{(c)}{=}H(V^n_1) \qquad \nonumber \\
&-H(K_1|\mathcal{C}) \geq n(H(V_1)-R_1), \nonumber
\end{align}}}where $(c)$ comes from the fact that $K_1$ is the sub-bin index of $V^n_1$, by substituting, we have
{\small{\begin{align}
&I(K_1;Z^n,\psi_1|\mathcal{C}) \leq n(R'_1+R_1-H(V_1|Z)) \nonumber \\
&+H(V^n_1|Z^n,\psi_1,K_1,\mathcal{C})\stackrel{(d)}{\leq} nR'_1+nR_1-nH(V_1|Z)+nH(V_1|Z) \nonumber \\
&-nR'_1-nR_1+n\epsilon=n\epsilon, \nonumber
\end{align}}}where $(d)$ follows by Lemma 22.3 in \cite{11} at which $H(V^n_1|Z^n,\psi_1,$ $K_1,\mathcal{C}) \leq nH(V_1|Z)-nR'_1-nR_1+n\epsilon$, if $R'_1+R_1 \leq H(V_1|Z)+\epsilon$. For the second term of \eqref{21},
{\small{ \begin{align}
&I(K_1;\psi_2|\psi_1,Z^n,\mc)=I(K_1,K_2,V^n_1,V^n_2;\psi_2|\psi_1,Z^n,\mc) \nonumber \\
&-I(V^n_1,V^n_2,K_2;\psi_2|K_1,\psi_1,Z^n,\mc) \stackrel{(a)}{=}I(V^n_1,V_2^n;\psi_2|\psi_1,Z^n,\mc) \nonumber \\
&-I(K_2;\psi_2|K_1,\psi_1,Z^n,\mc)-I(V^n_1,V_2^n;\psi_2|K_1,K_2,\psi_1,Z^n,\mc) \qquad \nonumber \\
&=I(V_1^n,V_2^n;K_1,K_2|\psi_1,Z^n,\mc)-H(V_1^n,V_2^n|\psi_1,\psi_2,Z^n,\mc) \nonumber \\
&+H(V_1^n,V_2^n|K_1,K_2,\psi_1,\psi_2,Z^n,\mc)+H(K_2|K_1,\psi_1,\psi_2,Z^n,\mc) \nonumber \\
&-H(K_2|K_1,\psi_1,Z^n,\mc) \stackrel{(b)}{\leq} nR_1+nR_2-nH(V_1,V_2|Z)+nR'_1 \nonumber \\
&+nR'_2+nH(V_1,V_2|Z)-nR'_1-nR_1-nR'_2-nR_2+n\epsilon=n\epsilon, \nonumber 
\end{align} }}$K_1$ and $K_2$ are the sub-bin indices of $V_1^n$ and $V_2^n$ the equality $I(K_1,K_2;\psi_2| \psi_1, V_1^n,V_2^n,Z^n, \mc)=0$ holds which implies $(a)$. $(b)$ can be deduced from Lemma 22.3 in \cite{11}. Finally, for the third term of \eqref{21} we have:
{ \small{ \begin{align}
&I(K_2;Z^n,\psi_1,\psi_2|K_1,\mc)=I(K_2,V^n_1,V^n_2;Z^n,\psi_1,\psi_2|K_1,\mc) \qquad \qquad \qquad \nonumber
\end{align}
\begin{align}
&-I(V^n_1,V^n_2;Z^n,\psi_1,\psi_2|K_1,K_2,\mc) \nonumber \\
&\stackrel{(a)}{=}I(V^n_1,V^n_2;Z^n,\psi_1,\psi_2|K_1,\mc) \nonumber \\
&-I(V^n_1,V^n_2;Z^n,\psi_1,\psi_2|K_1,K_2,\mc) \nonumber \\
&=H(V^n_1,V^n_2|K_1,\mc)-H(V^n_1,V^n_2|K_1,\psi_1,\psi_2,Z^n,\mc) \nonumber \\
&-H(V^n_1,V^n_2|K_1,K_2,\mc)+H(V^n_1,V^n_2|K_1,K_2,\psi_1,\psi_2,Z^n,\mc) \qquad \qquad \qquad \qquad \nonumber \\
&\stackrel{(b)}{=}H(K_2|K_1,\mc)-H(V^n_1,V^n_2|K_1,\psi_1,\psi_2,Z^n,\mc)\nonumber \\
&+H(V^n_1,V^n_2|K_1,K_2,\psi_1,\psi_2,Z^n,\mc), \nonumber
\end{align}}}as $K_2$ is the sub-bin index of $V_2^n$ the equality $I(K_2;Z^n,\psi_1, \\ \psi_2|K_1,V_1^n,V_2^n,\mc)=0$ holds which implies $(a)$. $(b)$ comes from the fact that $I(V_1^n,V_2^n;K_2|K_1,\mc)=H(K_2|K_1,\mc)$. Now
%\vspace{-2mm}
{\small{\begin{align}
&H(V^n_1,V^n_2|K_1,\psi_1,\psi_2,Z^n,\mc)=H(V^n_1,V^n_2,K_1,\psi_1,\psi_2|Z^n,\mc) \nonumber \\
&-H(K_1,\psi_1,\psi_2|Z^n,\mc) \stackrel{(c)}{=}H(V^n_1,V^n_2|Z^n,\mc) \nonumber \\
&-H(K_1,\psi_1,\psi_2|Z^n,\mc) \geq H(V^n_1,V^n_2|Z^n,\mc)-n(R'_1+R'_2+R_1)\nonumber \\
&=nH(V_1,V_2|Z)-n(R'_1+R'_2+R_1), \nonumber
\end{align}}}as $K_1$ is the sub-bin index of $V^n_1$ and also, $\psi_1$ and $\psi_2$ are the bin indices of $V_1^n$ and $V_2^n$, respectively, the equality $H(K_1,\psi_1,$ $\psi_2|V^n_1,V^n_2,Z^n,\mc)=0$ holds which implies $(c)$. Using Lemma 22.3 in \cite{11}, we have $H(V^n_1,V^n_2|K_1,K_2,$ $\psi_1,\psi_2,Z^n,\mc) \leq nH(V_1,V_2)-nR'_1-nR_1-nR'_2-nR_2-nI(V_1,V_2;Z)+n\epsilon$ if $R'_1+R'_2+R_1+R_2 \leq H(V_1,V_2|Z) + \epsilon$. Finally, by substituting we have $I(K_2;Z^n,\psi_1,\psi_2|K_1\mc) \leq n\epsilon$. Similarly, we can prove that the following constraints satisfy the secrecy conditions \eqref{11} and \eqref{12}:
\begin{align}
&R_1+R'_1 \leq H(V_1|V_2,Y_2), \nonumber \\
&R_1+R'_1 \leq H(V_2|V_1,Y_1). \label{24}
\end{align}

Collecting terms of \eqref{25}, \eqref{23} and \eqref{24}, and using Fourier-Motzkin elimination, we get the expressions \eqref{26}.
}

\begin{thebibliography}{99}
\bibitem{1}
C. E. Shannon, `` Communication theory of secrecy systems," \emph{Bell System Technical Journal}, vol. 28, pp. 656-715, 1949.
\bibitem{2}
R. Ahlswede and I. Csiszar, ``Common randomness in information theory and cryptography - Part I: Secret sharing,” \emph{IEEE Trans. Inf. Theory}, vol. 39, no. 4, pp. 1121-1132, Jul. 1993.
\bibitem{3}
U. M. Maurer, ``Secret key agreement by public discussion from common information," \emph{IEEE Trans. Inf. Theory}, vol. 39, no. 3, pp. 733-742, May 1993.
\bibitem{4}
A. A. Gohari and V. Anantharam, ``Information-theoretic key agreement of multiple Terminals - Part II: Channel model," \emph{IEEE Trans. Inf. Theory}, vol. 56, no. 8, pp. 3997-4010, Aug. 2010.
\bibitem{5}
A. Khisti, S. Diggavi, and G. Wornell, ``Secret-key generation using correlated sources and channels," \emph{IEEE~Trans.~Inf.~Theory}, vol.~58, no.~2, pp.~652-670, Feb. 2012.
\bibitem{6}
S. Salimi and M. Skoglund, ``Secret key agreement using correlated sources over the generalized multiple access channel," \emph{Arxiv preprint}, arXiv:~1204.2922v1, Apr. 2012.
\bibitem{7} A. Khisti, S. Diggavi, and G. Wornell, ``Secret key agreement with channel state information at the transmitter," \emph{IEEE Trans.~on Information Forensics and Security}, vol.~6, no.~3, pp.~672-681, Sep. 2011.
\bibitem{8} A. Khisti, ``Secret key agreement on wiretap channel with transmitter side information," in \emph{Proc.~European~Wireless~(EW)}, Lucca, Italy, pp.~802-809, Apr. 2010.
\bibitem{9} A. Khisti, S. Diggavi, and G. Wornell, ``Secret key agreement using asymmetry in channel state knowledge," in \emph{Proc.~Int.~Symp.~Inf.~Theory}, Seoul, Korea, pp.~2286-2290, Jun.-Jul. 2009.
\bibitem{10} D. Slepian and J. K. Wolf, ``Noiseless coding of correlated information sources," \emph{IEEE Trans.~Inf.~Theory}, vol.~19, no.~4, pp.~471-480, Jul. 1973.
\bibitem{11} A. El Gamal and Y. H. Kim, \emph{Network Information Theory}, 1st ed.~Cambridge University Press, 2011.
\end{thebibliography}
\end{document}